# Resonant leptogenesis with mild degeneracy

Naoyuki Haba[1], Osamu Seto[2], and Yuya Yamaguchi[1]

[1]*Department of Physics, Faculty of Science,*
*Hokkaido University, Sapporo 060-0810, Japan*
[2]*Department of Life Science and Technology,*
*Hokkai-Gakuen University, Sapporo 062-8605, Japan*

**Abstract**

Under the assumption of hierarchical right-handed neutrino masses, masses of right-handed neutrinos must be larger than $10^8$ GeV in the standard thermal leptogenesis scenario, while the mass can be reduced to around 5 TeV in a neutrinophilic two Higgs doublet model. On the other hand, resonant leptogenesis can work with the masses of TeV scale. However, necessary degeneracy between the lightest and the second-lightest right-handed neutrino masses means unnatural fine-tuning of the order of $10^9$. In this paper, we will investigate the resonant leptogenesis scenario in a neutrinophilic two Higgs doublet model. We will find the mass can be reduced to 2 TeV, and the degeneracy becomes much milder as of the order of $10^4$.

# 1 Introduction

In modern cosmology and particle physics, one of the important open problems is the origin of the baryon asymmetry in the Universe (BAU). Many kinds of models have been proposed in order to solve this problem; however, we do not know which model is true. The thermal leptogenesis scenario [1, 2] is an attractive candidate to explain the BAU, in which the simplest model introduces only right-handed heavy Majorana neutrinos in addition to the standard model (SM) [3]. Their $CP$-violating interactions make a lepton asymmetry during their out-of-thermal equilibrium, and, through the sphaleron process, a part of the lepton asymmetry turns into the baryon asymmetry. This simple procedure requires that the right-handed neutrino mass is larger than $10^8$ GeV [4, 5].

On the other hand, when the lightest and second-lightest right-handed neutrino masses are closely degenerate, the $CP$ asymmetry is enhanced by a self-energy of the right-handed neutrinos. Thanks to the large $CP$ asymmetry, so-called resonant leptogenesis [6] can explain the BAU even with the TeV-scale mass. We might have a chance to detect a TeV-scale new particle in collider experiments, such as the LHC and International Linear Collider [7, 8]. However, in the resonant leptogenesis, the mass degeneracy needs unnatural fine-tuning of $\mathcal{O}(10^9)$.

In order to explain the tiny neutrino masses in a novel approach, a new class of two Higgs doublet models (THDMs), so-called neutrinophilic THDMs, has been suggested [9]-[24]. The collider phenomenology of these models is studied in Refs. [15]-[18]. This model introduces an additional Higgs doublet, which has Yukawa interactions only with neutrinos. A vacuum expectation value (VEV) of the additional Higgs, $v_\nu$, is expected to have a much smaller energy scale compared to the SM Higgs doublet. This tiny VEV is an origin of the tiny neutrino masses without tiny couplings of neutrino Yukawa interactions. It plays a crucial role in a low-energy thermal leptogenesis as shown in Ref.[24], where it was shown that, around 5 TeV, mass of the right-handed neutrino can realize the suitable BAU. Note that from the measurement of flavor-changing neutral current, the neutrino Yukawa couplings should be smaller than $10^{-3}$; correspondingly, $v_\nu$ should be larger than 0.1 GeV for the right-handed neutrinos with TeV-scale mass.

In this paper, we will investigate the resonant leptogenesis scenario in a neutrinophilic THDM. We will find the masses of right-handed neutrinos can be reduced to 2 TeV, where the degeneracy becomes much milder as of order $\mathcal{O}(10^4)$.

## 2 Brief review of neutrinophilic THDM and leptogenesis

In this section, we review the neutrinophilic THDM and leptogenesis briefly.

### 2.1 Neutrinophilic THDM

Here, we review the neutrinophilic THDM [9], in which an additional Higgs doublet $\Phi_\nu$ and a discrete $Z_2$ parity are introduced. Their properties are assigned as Table 1. Under the discrete symmetry, Yukawa interactions are given by

$$-\mathcal{L}_{\text{yukawa}} = y^u \overline{Q_L} \Phi U_R + y^d \overline{Q_L} \tilde{\Phi} D_R + y^l \overline{L_L} \Phi E_R + y^\nu \overline{L_L} \Phi_\nu N + \frac{1}{2} M \overline{N^c} N + \text{h.c.} , \quad (1)$$

where $\tilde{\Phi} = i\sigma_2 \Phi^*$, $y$'s are Yukawa couplings, and $M$ shows masses of the right-handed neutrinos. Here, we omit the generation indices. Notice that new Higgs doublet gives only neutrino Dirac masses. Two different Higgs doublets, $\Phi$ and $\Phi_\nu$, are expected to have the nonzero VEVs, denoted by $v$ and $v_\nu$ (with $v > v_\nu$), respectively. The masses of light neutrinos are given by

$$m_{ij} = \sum_k \frac{y^\nu_{ik} v_\nu y^{\nu T}_{kj} v_\nu}{M_k} . \quad (2)$$

The neutrino Yukawa coupling, $y^\nu$, can be larger than the ordinary seesaw mechanism, which makes the leptogenesis scenarios work in the TeV scale, as we will show below.

| Fields | $Z_2$ parity | Lepton number |
|---|---|---|
| SM Higgs doublet, $\Phi$ | $+$ | 0 |
| New Higgs doublet, $\Phi_\nu$ | $-$ | 0 |
| Right-handed neutrinos, $N$ | $-$ | 1 |
| Others | $+$ | $\pm 1$: leptons, 0: quarks |

Table 1: Fields content of the neutrinophilic THDM.

### 2.2 Thermal leptogenesis

To explain the baryon asymmetry in the Universe, three conditions are required [25]: $B$-violating interaction, $C$ and $CP$ violation, and an out-of-thermal equilibrium process. Thermal leptogenesis easily satisfies the three conditions by introducing heavy right-handed Majorana neutrinos: an $L$-violating interaction of the right-handed Majorana neutrinos, $CP$ violating Yukawa interactions, and nonequilibrium decay due to the expansion of space. Through the sphaleron process, a part of the lepton asymmetry turns into the baryon asymmetry. We can estimate the $B-L$ asymmetry by solving the Boltzmann equations [4, 26]. In the simplest case, in which we take the only lightest right-handed neutrino into account, the Boltzmann equations are given by

$$\frac{dN_{N_1}}{dz} = -(D+S)(N_{N_1} - N_{N_1}^{\text{eq}}), \tag{3}$$

$$\frac{dN_{B-L}}{dz} = \varepsilon_1 D (N_{N_1} - N_{N_1}^{\text{eq}}) - W N_{B-L}, \tag{4}$$

where $z = M_1/T$, $M_1$ is the mass of the right-handed neutrino and $T$ is the temperature of the Universe. The number density of the right-handed neutrino $N_{N_1}$ and the amount of $B-L$ asymmetry $N_{B-L}$ are normalized in comoving volume, which contains one photon at temperatures $T \gg M_1$, so that the relativistic equilibrium $N_1$ number density is given by $N_{N_1}^{\text{eq}}(z \ll 1) = 3/4$. $D$ denotes the contribution of $N_1$ decays and inverse decays. $S$ denotes the contribution of $\Delta L = 1$ scatterings, mainly from top quark and gauge bosons. $W$ is the washout term, which contains the contribution of inverse decay, $\Delta L = 1$ scatterings, and $\Delta L = 2$ processes mediated by right-handed neutrinos. To keep our discussion conservative, we consider cases in which sufficient baryon asymmetry can be produced without relying on flavor effects, while the flavor effects can enhance the produced lepton asymmetry [27]-[30].

Using the Hubble expansion rate $H$ and interaction rates $\Gamma$'s, the contributions are written by $D = \Gamma_D/(Hz)$, $S = \Gamma_S/(Hz)$, and $W = \Gamma_W/(Hz)$, respectively. $D$ and $S$ depend on the effective neutrino mass [31], defined as

$$\widetilde{m}_1 \equiv \frac{(y^{\nu\dagger} y^{\nu})_{11} v^2}{M_1}, \tag{5}$$

where $y^\nu$ is the neutrino Yukawa coupling and $v = 174\,\text{GeV}$ is the Higgs vacuum expectation value. We define the decay parameter as

$$K_1 = \frac{\Gamma_{N_1}}{H(T=M_1)} = \frac{\widetilde{m}_1}{m_*}, \tag{6}$$

which represents whether $N_1$ decays are in equilibrium at $T = M_1$ or not. Here, $\Gamma_{N_1}$ is the $N_1$ decay width, and $m_*$ is the equilibrium neutrino mass, defined as

$$m_* \equiv \sqrt{\frac{8\pi^3 g_*}{90}} \frac{8\pi v^2}{M_{\text{Pl}}} \simeq 1.08 \times 10^{-3}\,\text{eV}, \tag{7}$$

where $g_*$ is the total number of degrees of freedom, and $M_{\text{Pl}} = 1.22 \times 10^{19}\,\text{GeV}$ is the Planck mass.

To solve the Boltzmann equations [(3) and (4)], we assume the initial number density of $N_{B-L}$ is zero. Then, we can get $N_{B-L}(z) = \frac{3}{4}\varepsilon_1 \kappa(z)$ [32], where $\varepsilon_1$ is the $CP$ asymmetry, and $\kappa$ is the efficiency factor [33]. Finally, considering a dilution factor that is calculated by the difference of $g_*$ and a conversion from lepton asymmetry to baryon asymmetry through sphaleron process [34], we obtain the final baryon asymmetry as

$$\eta_B \simeq \frac{86}{2387}\frac{28}{79} N_{B-L} \simeq 0.96 \times 10^{-2}\, \varepsilon_1\, \kappa_{\rm f}\,, \tag{8}$$

where $\kappa_{\rm f} = \kappa(\infty)$.

On the other hand, the present baryon-to-photon ratio of the number density has been measured by Wilkinson Microwave Anisotropy Probe [35] as

$$\eta_B^{CMB} = 6.19 \times 10^{-10}\,. \tag{9}$$

Equation (8) should be compared with this observed value.

## 2.3 Resonant leptogenesis

Next, we review the resonant leptogenesis [6]. $CP$ asymmetry is considerably enhanced through the mixing of two closely degenerate right-handed neutrinos $N_i$ ($i=1,2$). As a result, the lepton asymmetry produced by $N_i$ decays is enhanced, and the leptogenesis can work even by light $N_i$ with $\mathcal{O}(1)\,\text{TeV}$ masses.

The $CP$ asymmetry is given by [36, 37]

$$\begin{aligned}
\varepsilon_i &= \frac{\Gamma(N_i \to L\Phi) - \Gamma(N_i \to \overline{L}\Phi^*)}{\Gamma(N_i \to L\Phi) + \Gamma(N_i \to \overline{L}\Phi^*)} \\
&\simeq \frac{\operatorname{Im}(y^{\nu\dagger} y^\nu)_{ij}^2}{(y^{\nu\dagger} y^\nu)_{ii}\,(y^{\nu\dagger} y^\nu)_{jj}}\,\frac{\widetilde{m}_j M_j}{8\pi v^2}\,\frac{M_i M_j}{M_i^2 - M_j^2}\,,
\end{aligned} \tag{10}$$

where $i,j = 1,2$ ($i \neq j$) and the last factor expresses a mass degeneracy of two right-handed neutrinos. For $M_1 < M_2$, we define

$$d_N \equiv \frac{M_1 M_2}{M_2^2 - M_1^2}\,. \tag{11}$$

When $M_2 - M_1$ is small, $d_N$ is large; that is, $\varepsilon_i$ is large. Notice that small $M_2 - M_1$ means fine-tuning. The absolute value of the first factors composed of Yukawa couplings is less than unity, so we define them as $\sin\delta_1$ and $-\sin\delta_2$, respectively. Then, the $CP$ asymmetries are given by

$$\varepsilon_1 \simeq -\frac{\widetilde{m}_2 M_2}{8\pi v^2}\, d_N \sin\delta_1\,, \tag{12}$$

$$\varepsilon_2 \simeq -\frac{\widetilde{m}_1 M_1}{8\pi v^2}\, d_N \sin\delta_2\,. \tag{13}$$

The larger $d_N$ is, the larger both $\varepsilon_1$ and $\varepsilon_2$ are. For $M_1 \simeq M_2$, the difference between Eqs. (12) and (13) almost depends on the effective neutrino masses, $\widetilde{m}_1$ and $\widetilde{m}_2$.

Because of the large $CP$ asymmetries, the resonant leptogenesis can work with the right-handed neutrinos having TeV-scale masses, while the degeneracy needs terrible fine-tuning of $\mathcal{O}(10^9)$ in an ordinary resonant leptogenesis. It is a disadvantage with the model.

## 3 Resonant leptogenesis in the neutrinophilic THDM

Now, let us investigate the leptogenesis in the neutrinophilic THDM. In the neutrinophilic THDM, the Boltzmann equation for the lepton asymmetry $L \equiv l - \bar{l}$ is given by

$$\begin{aligned}
&\dot{n}_L + 3Hn_L \\
= \ & \gamma(N \to l\Phi_\nu) - \gamma(N \to \bar{l}\Phi_\nu^*) \\
& -\{\gamma(l\Phi_\nu \to N) - \gamma(\bar{l}\Phi_\nu^* \to N)\} \quad : \text{decays and inverse decays} \\
& -\gamma(lA \to N\Phi_\nu) + \gamma(\bar{l}A \to N\Phi_\nu^*) - \gamma(Nl \to A\Phi_\nu) + \gamma(N\bar{l} \to A\Phi_\nu^*) \\
& -\gamma(l\Phi_\nu \to NA) + \gamma(\bar{l}\Phi_\nu^* \to NA) \quad : \text{s- and t-channel } \Delta L = 1 \text{ scatterings} \\
& +\gamma(\bar{l}\bar{l} \to \Phi_\nu^*\Phi_\nu^*) - \gamma(ll \to \Phi_\nu\Phi_\nu) \\
& +2\{\gamma'(\bar{l}\Phi_\nu^* \to l\Phi_\nu) - \gamma'(l\Phi_\nu \to \bar{l}\Phi_\nu^*)\} \quad : \text{s- and t-channel } \Delta L = 2 \text{ scatterings} \\
= \ & \varepsilon \Gamma_D (n_N - n_N^{\text{eq}}) - \Gamma_W n_L,
\end{aligned} \quad (14)$$

where we omit the generation indices. $\Phi_\nu$ and $A$ denote the neutrinophilic Higgs bosons and gauge bosons, respectively. $\gamma$ terms describe the change of the number densities due to the corresponding interactions. Here $\gamma'$ terms are the same as $\gamma$ terms up to additional subtraction of the real right-handed neutrinos mediated scattering processes. The washout rate is given by

$$\Gamma_W = \frac{1}{2}\frac{n_N^{\text{eq}}}{n_l^{\text{eq}}}\Gamma_N + \frac{n_N}{n_N^{\text{eq}}}\Gamma_{\Delta L=1,\text{s}} + 2\Gamma_{\Delta L=1,\text{t}} + 2\Gamma_{\Delta L=2}. \quad (15)$$

If the initial $n_B$ is zero, the solution of the equation $n_L$ is equal to $-n_{B-L}$, and Eq. (14) reduces to Eq. (4).

Note that we take the maximal contribution of the $\Delta L = 2$ processes as a upper bound of the washout rate, Eq. (15). For $T < M_1$, the decoupling condition is given by [24]

$$\sum_i \left( \sum_j \frac{y_{ij}^\nu y_{ji}^{\nu\dagger} v_\nu^2}{M_j} \right)^2 < 32\pi^3 \zeta(3) \sqrt{\frac{\pi^2 g_*}{90}} \frac{v_\nu^4}{TM_P}. \quad (16)$$

Thus, the $\Delta L = 2$ washout processes are more significant for lower $v_\nu$. The above inequality gives the lower bound on $v_\nu$ in order to avoid too strong a washout. We will use this bound for the numerical results.

## 4 Numerical analyses

Before solving the Boltzmann equations, we recall the condition of the out-of-equilibrium decay. Using the decay parameter, Eq. (6), the condition $K_1 < 1$ becomes

$$\widetilde{m}_1 < m_* \left(\frac{v_\nu}{v}\right)^2 . \tag{17}$$

Notice that there is an additional factor, $(v_\nu/v)^2$, in the neutrinophilic Higgs model. When we concentrate on the case $v_\nu \ll v$, the effective neutrino masses are given by

$$\widetilde{m}_1 \simeq 0 \text{ eV}, \tag{18}$$
$$\widetilde{m}_2 \simeq \sqrt{\Delta m_{21}^2} \equiv m_{\text{sol}} \simeq 8.6 \times 10^{-3} \text{ eV} . \tag{19}$$

Then, the $CP$ asymmetry Eqs. (12) and (13) are given by

$$\varepsilon_1 \sim -\frac{\widetilde{m}_2 M_2}{8\pi v_\nu^2} d_N \sin\delta_1 = -\frac{y_\nu^2}{8\pi} d_N \sin\delta_1 , \tag{20}$$
$$\varepsilon_2 \sim 0 , \tag{21}$$

where $y_\nu^2$ denotes $(y^{\nu\dagger} y^\nu)_{22}$. Using these parameters, we solve the Boltzmann equations. We will show the results of numerical calculations below.

We consider a scenario with two nearly degenerate right-handed neutrinos $N_{1,2}$, for which the masses are a few TeV, and neglected $N_3$, for which the mass is much heavier than $N_{1,2}$. We will show the following three dependences: (1) the neutrino Yukawa coupling $y_\nu$ dependence of the final baryon asymmetry $\eta_B$, (2) the neutrinophilic Higgs VEV $v_\nu$ dependence of the $CP$ asymmetry $\varepsilon_1$ and the minimum degeneracy of right-handed neutrino masses $d_{N\min}$, and (3) the right-handed neutrino mass $M_1$ dependence of $\eta_B$ and $d_N$.

At first, let us show the solution of the Boltzmann equations. Figure 1 shows the evolution of the lepton asymmetry for $M_1 = 2\,\text{TeV}$ without a sphaleron effect. The generation of the lepton asymmetry is completed at around $z_{\text{fin}} \simeq 20$. The sphaleron process ceases at $z_{\text{sph}} = M_1/T_{\text{sph}} \simeq 20$, because the electroweak symmetry breaking takes place around 100 GeV, and we may consider $T_{\text{sph}} \simeq 100\,\text{GeV}$. If $z_{\text{fin}}$ is smaller than $z_{\text{sph}}$, before the sphaleron process ceases, $N_{B-L}$ is frozen out, which is shown as the plateau in Fig. 1. Note that a low-energy leptogenesis such as $\mathcal{O}(100)\,\text{GeV}$ includes two different types of uncertainties about the sphaleron process. One is that the relation $z_{\text{fin}} \leq z_{\text{sph}}$ is usually not satisfied for small $M_1$. So, $N_{B-L}$ is not frozen out while the sphaleron process is active, which leads the final production of the lepton asymmetry to be unclear. The other is about the condition of whether the sphaleron process is really in thermal equilibrium, which is roughly expressed by $H(T_{\text{sph}}) \lesssim \Gamma_{N_1} = K_1 H(T = M_1)$ and rewritten as

$$K_1 \gtrsim \frac{H(T_{\text{sph}})}{H(T = M_1)} = \left(\frac{T_{\text{sph}}}{M_1}\right)^2. \tag{22}$$

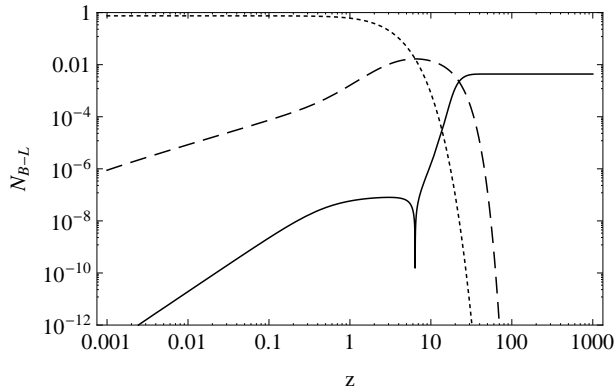

Figure 1: Time evolution of $N_{B-L}$ with $\varepsilon_1 = -1$, $M_1 = 2\,\text{TeV}$, $y_\nu = 10^{-4}$, and $K_1 = 10^{-2}$. The dashed, dotted, and solid lines correspond to $N_1, N_2$, and $N_{B-L}$, respectively.

We find $K_1 \gtrsim 2.5 \times 10^{-3}$ for $M_1 = 2\,\text{TeV}$. It means that, for $K_1 \gtrsim 2.5 \times 10^{-3}$, the sphaleron process completely works, and the lepton asymmetry turns into baryon asymmetry according to the relation $N_B = (28/79) N_{B-L}$. On the other hand, for $K_1 \lesssim 2.5 \times 10^{-3}$, the validity of using the relation is unclear. In this paper, we do not carefully treat these uncertainties so much, while it could be a crucial point for low-scale thermal leptogenesis. Notice that, in the $M_1 = 2\,\text{TeV}$ case, the resultant lepton asymmetry at $K_1 = 10^{-2}$ is not affected by these uncertainties.

Figures 2(a) and 2(b) show the $y_\nu$ dependence of the final baryon asymmetry, $\eta_B(z \gg 1)$ with $\varepsilon_1 = -1$. Here, $\eta_B(z \gg 1)$ means that we consider the lepton asymmetry, which is frozen out. As discussed above, when $M_1$ is small as $\mathcal{O}(100)\,\text{GeV}$, $\eta_B(z \gg 1)$ is changeful. To explain the BAU, at least $\eta_B$ with $\varepsilon_1 = -1$ should be larger than $\eta_B^{CMB}$ because $\eta_B$ is almost proportional to $\varepsilon_1$, and the absolute value of $\varepsilon_1$ is less than unity. So, if $\eta_B$ with $\varepsilon_1 = -1$ is larger than $\eta_B^{CMB}$, the observed baryon asymmetry can be always reproduced by taking $\varepsilon_1$ to the smaller value. In Fig. 2(a), as $y_\nu$ is large, the baryon asymmetry extremely decreases between $K_1 = 0.01$ and $K_1 = 0.1$ and survives some constant value for $K_1 > 1$,

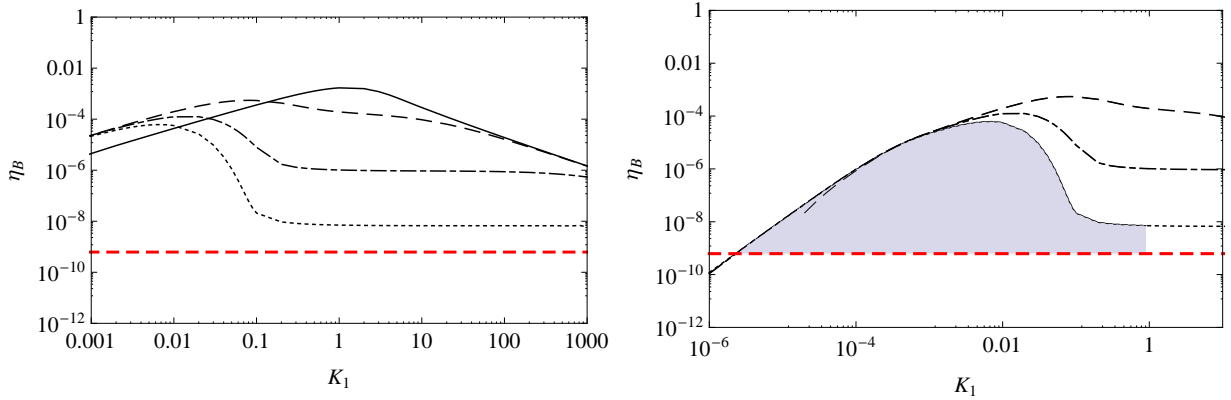

(a) $K_1$ dependence of $\eta_B$ with $10^{-3} \leq K_1 \leq 10^3$

(b) $K_1$ dependence of $\eta_B$ with $10^{-6} \leq K_1 \leq 10$

Figure 2: $\eta_B(z \gg 1)$ with $\varepsilon_1 = -1$ and $M_1 = 2\,\text{TeV}$. The solid, dashed, dotted-dashed, and dotted lines correspond to $y_\nu = 10^{-7}, 10^{-6}, 10^{-5}$, and $10^{-4}$, respectively. Here, (a) contains all, while (b) does not contain $y_\nu = 10^{-7}$. The red thick-dashed line shows the observed value, $\eta_B^{CMB} = 6.19 \times 10^{-10}$.

except $y_\nu = 10^{-7}$. When $y_\nu$ is smaller than $10^{-7}$, the results is nearly the same as an ordinary resonant leptogenesis.

The behavior shown in Fig. 2(a) is caused by an $N_2$ washout effect, which becomes stronger as $y_\nu$ is large. Here, we have to remind the reader that we use the condition of the out-of-equilibrium decay, $K_1 < 1$, where $\varepsilon_2 \sim 0$. For a non-negligible $\varepsilon_2$, $\eta_B$ at $K_1 = 1$ is larger than the value shown in Fig. 2(a) by a $\mathcal{O}(1)$ factor. Figure 2(b) shows the same lines (without the line of $y_\nu = 10^{-7}$) as Fig. 2(a) in the lower region of $K_1$, where all lines represent that enough baryon asymmetry is produced at least for $K_1 \gtrsim 2 \times 10^{-6}$. We find that this bound corresponds to $10^{-15}\,\text{eV} < \widetilde{m}_1 < 10^{-8}\,\text{eV}$ for $y_\nu = 10^{-4}$ (illustrated by shaded region), for which the upper bound is determined by $K_1 < 1$. This small $\widetilde{m}_1$ means that the degenerate mass spectrum of active neutrinos is disfavored, which is the same as the ordinary resonant leptogenesis.

Figure 3(a) shows $\varepsilon_1$, which agrees with the BAU, and Fig. 3(b) shows the minimum of $d_N$ defined as Eq. (11) for the neutrinophilic Higgs VEV, $v_\nu$. From Eq. (12), we obtain the inequality

$$d_N \gtrsim -\varepsilon_1 \frac{8\pi v_\nu^2}{\widetilde{m}_2 M_2}. \tag{23}$$

Fitting the value of $\varepsilon_1$, that is, taking $\eta_B(z \gg 1) = \eta_B^{CMB}$, we obtain the minimum of $d_N$, which is denoted by $d_{N\min}$. In Fig. 3(a), $\varepsilon_1$ is almost proportional to the $v_\nu^{-2}$ for $K_1 \gtrsim 0.1$. In Fig. 3(b), $d_{N\min}$ is nearly constant around $10^8$ for $K_1 \gtrsim 0.1$. Here, $d_N = 10^9$ is a typical value of the degeneracy in the ordinary resonant leptogenesis. For $K_1 \lesssim 0.01$, $\varepsilon_1$ is almost constant; correspondingly, $d_{N\min}$ is almost proportional to the $v_\nu^2$.

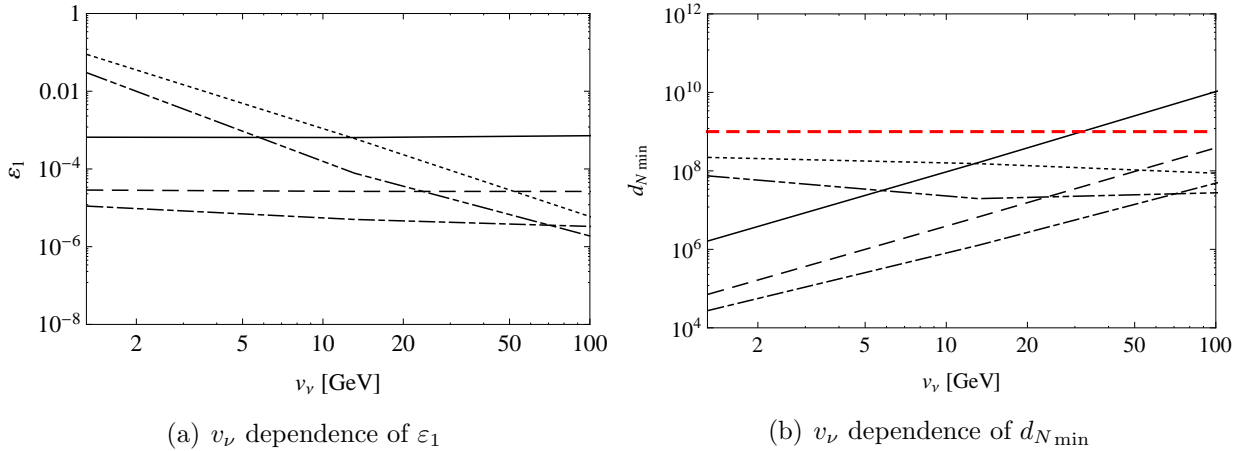

(a) $v_\nu$ dependence of $\varepsilon_1$

(b) $v_\nu$ dependence of $d_{N\min}$

Figure 3: $v_\nu$ dependence of (a) $\varepsilon_1$ and (b) $d_{N\min}$, with $M_1 = 2\,\text{TeV}$. $\varepsilon_1$ is obtained by $\eta_B(z \gg 1) = \eta_B^{CMB}$. The numerical results are shown by the solid, dashed, dotted-dashed, dotted-dotted-dashed, and dotted lines, which correspond to $K_1 = 10^{-4}, 10^{-3}, 10^{-2}, 10^{-1}$, and 1, respectively. The red thick-dashed line shows the degeneracy of an ordinary resonant leptogenesis such as $d_N = 10^9$.

Note that the relation $d_N \propto v_\nu^2$ is a unique behavior in the neutrinophilic THDM because $v_\nu$ is the unique parameter in the model, while the VEV, which provides the Dirac neutrino mass, is fixed by the SM Higgs VEV. Thanks to the behavior, we obtain the minimum of the mass degeneracy such as $\mathcal{O}(10^4)$ for small $v_\nu$. When $v_\nu$ is smaller than $1\,\text{GeV}$ (correspondingly, $y_\nu$ is larger than $10^{-4}$), we can obtain the smaller value of $d_N$. But, from the measurement of the flavor-changing neutral current, $y_\nu > 10^{-3}$ ($v_\nu < 0.1\,\text{GeV}$) has been ruled out. And, it was shown that, when we include the contribution of the $\Delta L = 2$ processes in the Boltzmann equations, $v_\nu < 0.3\,\text{GeV}$ is washed out for $M_1 = 2\,\text{TeV}$ [24]. So, we need not consider the small $v_\nu$ such as $0.1\,\text{GeV}$ (large $y_\nu$ such as $10^{-3}$).

Finally, we check the dependence on the right-handed neutrino mass. The $M_1$ dependence of $\eta_B$ is shown in Fig. 4. In the figure, for $K_1 \lesssim 0.01$, there is almost no difference, while for $K_1 \gtrsim 0.1$, there is much difference. For each of them, the mass dependence of the minimum degeneracy are shown in Figs. 5(a) and 5(b). Notice that $d_{N\min}$ is almost constant for $K_1 = 0.01$ in Fig. 5(a), while $d_{N\min}$ is nearly proportional to $M_1$ for $K_1 = 0.1$ in Fig. 5(b). In the latter case, $d_{N\min}$ becomes smaller as $M_1$ becomes large. We obtain the lowest value of the degeneracy from the former case due to the small $\varepsilon_1$.

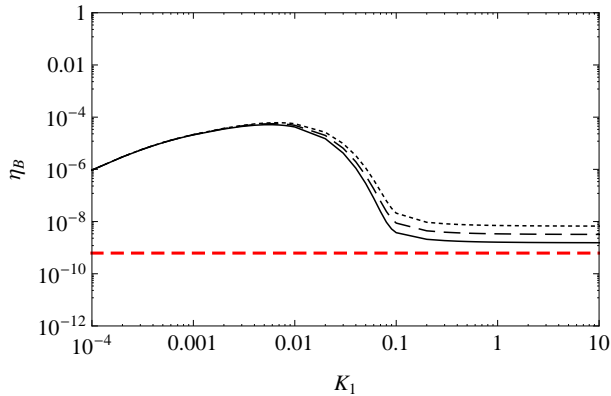

Figure 4: $M_1$ dependence of $\eta_B$. Here $\eta_B$ is estimated with $\varepsilon_1 = -1$ and $y_\nu = 10^{-4}$. The solid, dashed and dotted lines correspond to $M_1 = 500\,\text{GeV}$, $1\,\text{TeV}$, and $2\,\text{TeV}$, respectively. The red thick-dashed line shows the observed value, $\eta_B^{CMB} = 6.19 \times 10^{-10}$.

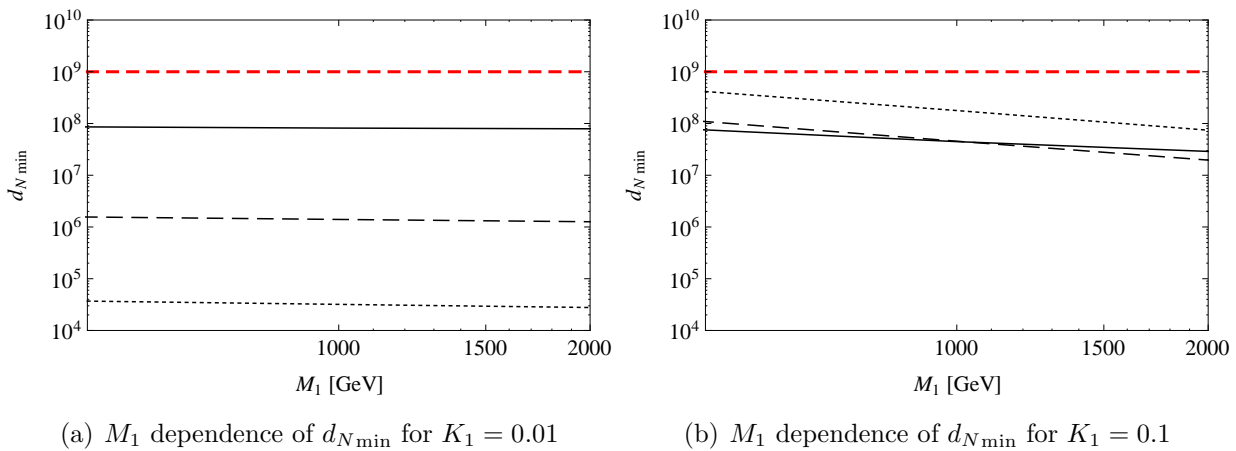

(a) $M_1$ dependence of $d_{N\min}$ for $K_1 = 0.01$

(b) $M_1$ dependence of $d_{N\min}$ for $K_1 = 0.1$

Figure 5: $M_1$ dependence of $d_{N\min}$ for (a) $K_1 = 10^{-2}$ and (b) $K_1 = 10^{-1}$. The solid, dashed, and dotted lines correspond to $y_\nu = 10^{-6}, 10^{-5}$, and $10^{-4}$, respectively. The red thick-dashed line shows the degeneracy of an ordinal resonant leptogenesis such as $d_N = 10^9$.

As a summary, we show rough values of $d_{N\min}$ and $\widetilde{m}_1$ for each $y_\nu$ in Table. 2. The bounds for $\widetilde{m}_1$ are given by the TeV-scale mass of right-handed neutrinos. Note that the $y_\nu = 10^{-7}$ case almost corresponds to the ordinal resonant leptogenesis scenario.

The observed neutrino mass differences and mixing angles can be reproduced by taking suitable masses and Yukawa couplings of the right-handed neutrinos. We can easily take its parameter set, which is not constrained by the LHC experiment. It is because right-handed neutrinos are gauge singlets and have too small a mixing with left-handed neutrinos to be observed by a direct detection. Actually, the charged Higgs boson, which is composed almost of neutrinophilic Higgs, could be observed at the LHC in the particular set of parameters, such as a case in which the right-handed neutrinos are heavier than the charged Higgs boson [18]. But it depends on the charged Higgs mass, which is beyond the contents of our paper.

|  | Order of $d_{N\min}$ | Order of bound for $\widetilde{m}_1$ |
|---|---|---|
| $y_\nu = 10^{-4}$ | $10^4$ | $10^{-15}\,\text{eV} \sim 10^{-8}\,\text{eV}$ |
| $y_\nu = 10^{-5}$ | $10^6$ | $10^{-15}\,\text{eV} \sim 10^{-6}\,\text{eV}$ |
| $y_\nu = 10^{-6}$ | $10^8$ | $10^{-15}\,\text{eV} \sim 10^{-4}\,\text{eV}$ |
| $y_\nu = 10^{-7}$ | $10^9$ | $10^{-15}\,\text{eV} \sim 10^{-2}\,\text{eV}$ |

Table 2: Values of $d_{N\min}$ and $\widetilde{m}_1$ for each $y_\nu$.

## 5 Summary and discussions

We have studied the resonant leptogenesis in the neutrinophilic THDM. Although usual thermal leptogenesis requires the right-handed neutrino mass to be larger than $10^8\,\text{GeV}$, the neutrinophilic THDM can reduces the mass to around $5\,\text{TeV}$ [24]. On the other hand, resonant leptogenesis works with the masses of $\mathcal{O}(1)\,\text{TeV}$; however, the degeneracy between the lightest and the second-lightest right-handed neutrino masses requires unnatural fine-tuning of $\mathcal{O}(10^9)$. In this paper, we have shown the resonant leptogenesis works with the right-handed neutrino masses of $2\,\text{TeV}$ in the neutrinophilic THDM, where the fine-tuning of the mass degeneracy can be much smaller as $\mathcal{O}(10^4)$.

Finally, we comment on how small we can take the masses of right-handed neutrinos. If we consider the low-energy thermal leptogenesis of $\mathcal{O}(100)\,\text{GeV}$, we have to mind two conditions about a sphaleron process: (1) whether or not the lepton asymmetry is frozen out before the sphaleron process is finished and (2) whether or not the sphaleron process is in thermal equilibrium. If both answers are positive, we do not have a problem. If not, we must treat an uncertainty, and when the right-handed neutrino masses are of $\mathcal{O}(100)\,\text{GeV}$, the model usually contains this uncertainty. Thus, if we consider the thermal leptogenesis of $\mathcal{O}(100)\,\text{GeV}$, we must estimate carefully. This is why we took the result that the right-handed neutrino masses are $2\,\text{TeV}$.

# Acknowledgments

We would like to thank R. Takahashi for valuable discussions. This work is partially supported by the Scientific Grant by the Ministry of Education and Science, Grants No. 00293803, No. 20244028, No. 21244036, and No. 23340070, and by the SUHARA Memorial Foundation.